\documentclass[twocolumn,showpacs,preprintnumbers,amsmath,amssymb,prl,a4paper,superscriptaddress]{revtex4}
\usepackage{amssymb,amsmath,amsbsy}
\usepackage{mathrsfs}
\usepackage{graphicx}
\usepackage{dcolumn}
\usepackage{bm}
\usepackage{verbatim}
\usepackage{subfigure}
\usepackage{color}

\newcommand{\tmyag}{Tm$^{3+}$:YAG}
\definecolor{green2}{rgb}{0,0.5,0}
\begin{document}
\title{Experimental realization of light with time separated correlations by rephasing amplified spontaneous emission}

\author{Patrick M. Ledingham}
\altaffiliation{Present address: ICFO-Institut de Ciencies Fotoniques, Av. Carl Friedrich Gauss 3, 08860 Castelldefels (Barcelona), Spain}
\affiliation{Jack Dodd Centre for Photonics and Ultra-Cold Atoms, Department of Physics, University of Otago, Dunedin, New Zealand.}

\author{William R. Naylor}
\altaffiliation{Present address: Department of Physics, Norwegian University of Science and Technology, N-7491 Trondheim, Norway}
\affiliation{Jack Dodd Centre for Photonics and Ultra-Cold Atoms, Department of Physics, University of Otago, Dunedin, New Zealand.}

\author{Jevon J. Longdell}
\email{jevon.longdell@otago.ac.nz}
\affiliation{Jack Dodd Centre for Photonics and Ultra-Cold Atoms, Department of Physics, University of Otago, Dunedin, New Zealand.}

\date{\today}
\begin{abstract}
Amplified spontaneous emission is a common noise source in active optical systems, it is generally seen as being an incoherent process. Here we excite an ensemble of rare earth ion dopants in a solid with a $\pi$-pulse, resulting in amplified spontaneous emission. The application of a second $\pi$-pulse leads to a coherent echo of the amplified spontaneous emission that is correlated in both amplitude and phase. For small optical thicknesses, we see evidence that the amplified spontaneous emission and its echo are entangled. 
\end{abstract}
\maketitle
Amplified spontaneous emission (ASE) \cite{alle73} is a ubiquitous phenomena that produces low-temporal coherence light in optical amplifiers. As well as being an unwanted noise process in optical amplifiers \cite{henr86}, and a source of inefficiency in lasers \cite{barn99}, it forms the basis of many useful high brightness, broadband light sources \cite{wyso94}. Recent theoretical work showed that applying rephasing pulses to atoms with long coherence times produces an `echo' of  ASE \cite{ledi10}. It was also shown, theoretically, that the photon count correlations between the rephased amplified spontaneous emission (RASE) and the ASE violate the classical Cauchy-Schwartz inequality for small optical depths.

RASE is attractive in that streams of time separated entangled photons can be created. This is interesting in the context of scaleable, long distance quantum information networking and communication, which requires a quantum repeater \cite{Briegel1998}. At its heart, a quantum repeater is a source of entangled photons that are separated in time. This way entanglement swapping can entangle two remote stations even though the photons coming from these stations may arrive at different times.
One proposal to realize an elementary link of a quantum repeater using atomic ensembles and linear optics is the Duan-Lukin-Cirac-Zoller (DLCZ) protocol \cite{dlcz}.
The DLCZ protocol uses atoms in a lambda configuration. It relies on the fact that every excitation of the spin-wave is accompanied by the absorption of a photon from one of the two coupled optical fields and emission into the other. In the write process the photon is absorbed from the coherent driving field and emitted into the signal field. As a result of this there is a one to one correspondence between the excitations created in the spin-wave and in the signal field. The read process is followed some time later by the write process, where the roles of the two fields are reversed. This time there is a one to one correspondence between the excitations removed from the spin-wave and photons emitted into the signal field. This results in a photon pair source with the photons separated in time. Among the recent achievements in the implementation of the DLCZ protocol are long term storage in a system that produces telecommunication wavelength write photons \cite{Dudin10} and the entanglement of four ensembles \cite{choi10}.

RASE has strong parallels to the DLCZ protocol, especially when the upper level of the lambda system can be adiabatically eliminated, meaning the lambda systems can be treated effectively as two level atoms coupled to the signal fields, with the coherent driving field determining the strength of the coupling \cite{ledi10}. 

In RASE, an inhomogeneously broadened collection of two level atoms are initially placed in the excited state, resulting in gain and therefore ASE. For each ASE photon emitted one atom is transferred to the ground state, but like in the case of the DLCZ protocol this de-excitation is coherently spread among the atoms as a collective excitation. A $\pi$-pulse then inverts the ensemble, placing the majority of the atoms in the ground state. The coherence in the atoms produced with the ASE is rephased resulting in collective emission of the RASE. This is different to DLCZ where swapping the control and signal fields means that the effective two level atoms have their ground and excited states reversed. A strong advantage that this gives RASE is that inhomogeneous broadening, rather than limiting the storage, means that the process is temporally multimode with associated improvements for quantum repeater operations \cite{simo07}. Another way proposed to make a DLCZ-like protocol that is multimode involves structuring the optical inhomogeneous broadening \cite{seka11} in a similar way to atomic frequency comb memories \cite{afze09}. The advantage that RASE has over such approaches is that it does not require tailoring the inhomogeneous absorption profile with spectral holeburning. Spectral holeburning with high contrast and fine frequency selectivity to date requires the use of hyperfine structure in non-Kramers ions and the hyperfine structure then limits bandwidths to $\lesssim100\,$MHz. 
In the case of RASE in \tmyag\ considered here, the inhomogeneous broadening is approximately 30~GHz. In principle this means very large bandwidth operation is possible, and in our results the bandwidth is only limited by the bandwidth of the exciting and rephasing $\pi$-pulses. 

Another route to a quantum repeater is to use a more standard parametric downconversion source of photon pairs and then store one of these in a quantum memory based on rare earth ion dopants. This has recently been demonstrated \cite{Clausen2011b,Saglamyurek2011b}. This approach holds great promise because of the storage time \cite{stopped}, bandwidths \cite{usma10}, and efficiencies \cite{hedg10,Chaneliere2010b} that have recently been achieved. Using RASE rather than downconversion for the sources in such experiments would greatly simplify the experiments, naturally providing sources with the appropriate brightness, wavelength and bandwidths.

Here we report on correlated, time separated, optical fields using RASE in cooled thulium doped (0.1\%) yttrium aluminum garnet (\tmyag).  
The thulium ions formed the required inhomogeneously broadened ensemble of two level atoms via their $^3$H$_6 \leftrightarrow ^3$H$_4$ transitions. Shot-noise limited balanced heterodyne detection is used to characterize the ASE and the RASE by measuring variances of the light quadratures $\hat{x}$ and $\hat{p}$. The experimental setup is shown in \cite{supp}.
The benefit of using \tmyag\  is that, at zero magnetic field, it lacks a long term spectral holeburning mechanism. A long term spectral holeburning mechanism is undesirable because it generally leads to inadvertent structure being prepared in the inhomogeneous absorption profile. This structure, even with very low absorption contrast along with the driving pulses would have lead to optical free-induction decays (FIDs) which mask the ASE and RASE fields. At zero magnetic field there is no structure for the ground and excited states we are using. The longest lived spectral holeburning mechanism is due to a metastable electronic level and has a lifetime of $\sim$~10~ms \cite{Macfarlane93}. The 10~Hz repetition rate used for the experiment ensures that this level was effectively emptied between shots. Another way to avoid the random FIDs associated with spectral holeburning is the four level scheme of \cite{beav11}.

\begin{figure}[t!]
\subfigure[]{\includegraphics[width=0.45\textwidth]{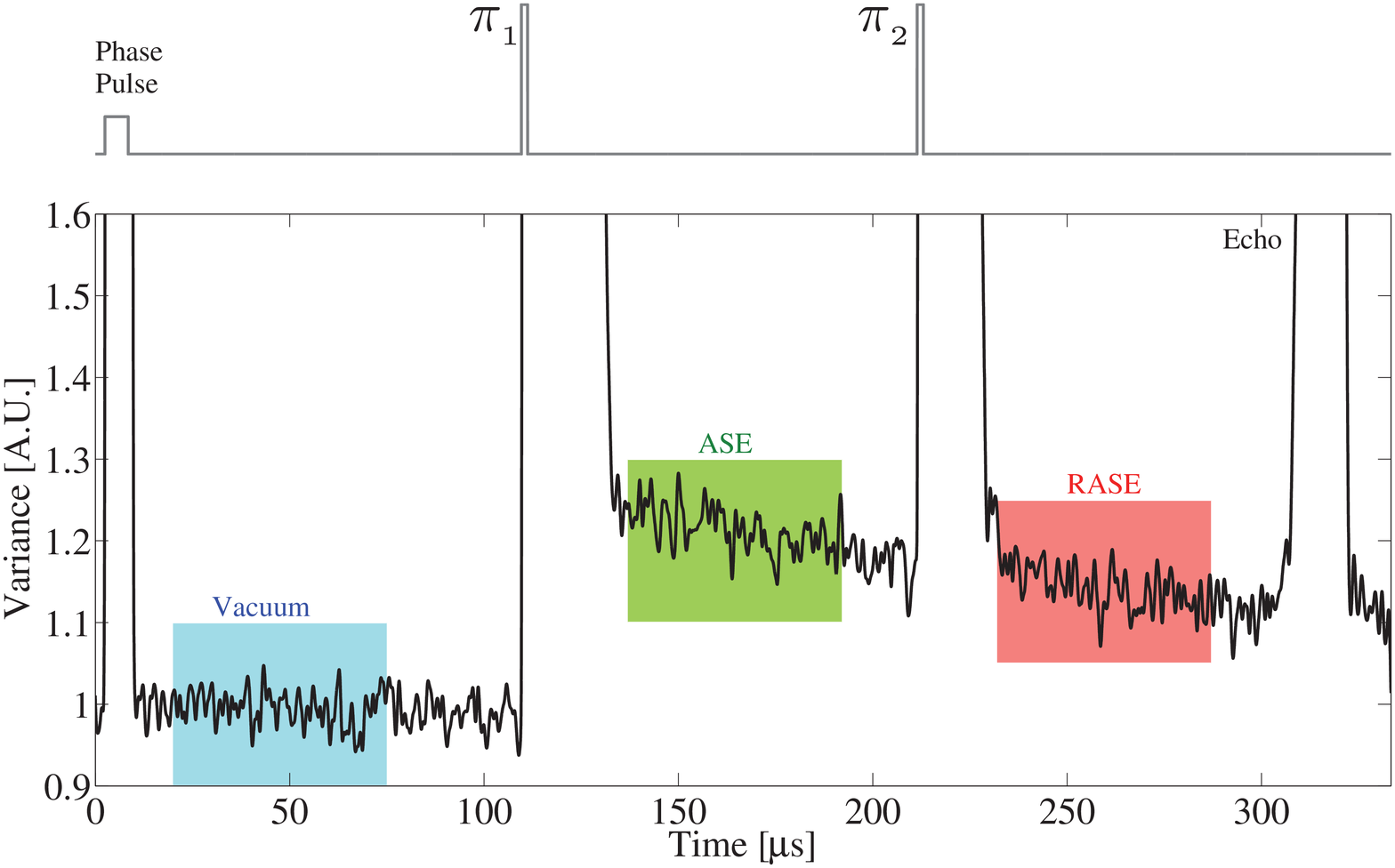}}
\subfigure[]{\includegraphics[width=0.48\textwidth]{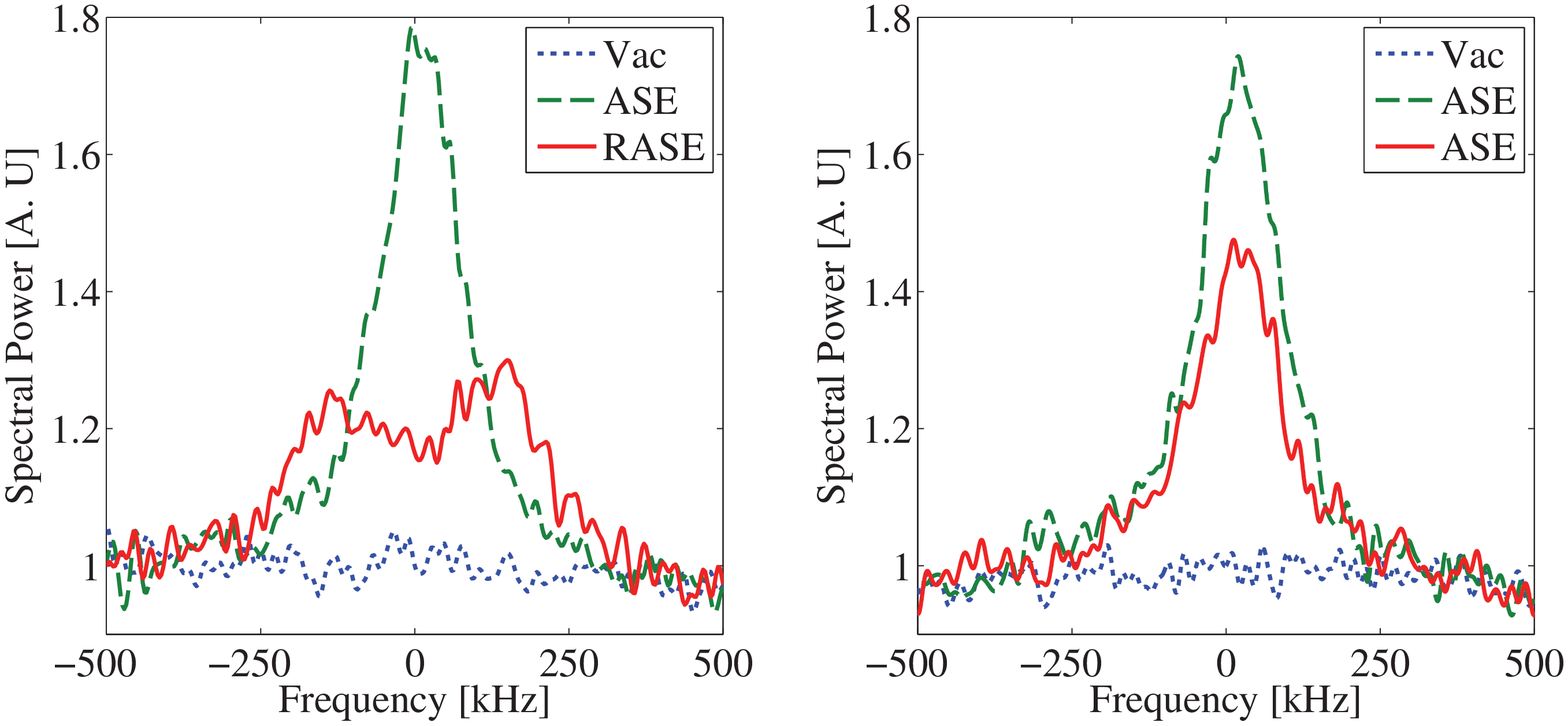}}
\caption{(Color online) (a) The pulse sequence used along with the sum of the variances of $\hat{x}$ and $\hat{p}$ as a function of time for an optically thick sample. Within experimental uncertainty, both of the ASE and RASE signals were phase independent and the variances of $x$ and $p$ equal.  The pulse marked 'Echo' is the inadvertent echo caused by the two $\pi$-pulses.  (b) The spectral power of the three regions for two different cases. Left panel: Spectra for the pulse sequence seen in (a). Right panel:  Spectra for the same time windows but with $\pi_2$ removed. 
The data are normalized to give the vacuum region a variance of one.}
\label{fig:RASEvsTime}
\end{figure}

Figure~\ref{fig:RASEvsTime}(a) shows the pulse sequence used in these experiments and the variance of the recorded heterodyne signals as a function of time. In order to improve the signal to noise ratio these experiments were carried out on a physically and optically thick sample (20~mm, $\alpha l = 3.2$).  The first pulse is used to measure the phase of the interferometer used for the heterodyne detection. It contained enough photons to make an accurate measurement of the phase but was orders of magnitude smaller than a $\pi$-pulse. It's effect on the ions can be ignored because of its small size and because any excitation produced is not rephased during the ASE and RASE windows.

 The~1.6~$\mu$s $\pi_1$-pulse inverts the part of the inhomogeneous line resulting gain and ASE. The ASE can be seen in Fig.~\ref{fig:RASEvsTime}(a) as an increase in the variance of the heterodyne signals after $\pi_1$. The ASE decay time is 378~$\mu$s, an order of magnitude larger than $T_2$ (13~$\mu$s, measured with two pulse photon echoes). This rules out the possibility that this is  FIDs which would decay at a timescale given by the coherence time.

Figure~\ref{fig:RASEvsTime}(b) shows the shows the spectra of the complex valued heterodyne signals  ($z(t) = x(t)+ip(t)$) for the time intervals indicated in Fig.~\ref{fig:RASEvsTime}(a). We can see that the spectrum of the ASE light has a 150~kHz wide peak, comparable to the Fourier width of the 1.6~$\mu$s $\pi_1$-pulse. 
One could expect a significant drop in the variance after the second $\pi$-pulse. This is because at this point the majority of ions are still excited and $\pi_2$ should return these back to the ground state. This is not seen clearly in the plot of the variance (Fig.~\ref{fig:RASEvsTime}(a)) because of the contributions of off-resonant ions which do not see accurate $\pi$-pulses. This difference in behaviour of the off-resonant and on-resonant ions can be seen in Fig.~\ref{fig:RASEvsTime}(b), where it is seen that $\pi_2$ causes the light from ions resonant with the driving laser to decrease significantly, whereas the signal from some off-resonant ions increases.

%

\begin{figure}[t!]
\subfigure[]{\includegraphics[width=0.21\textwidth]{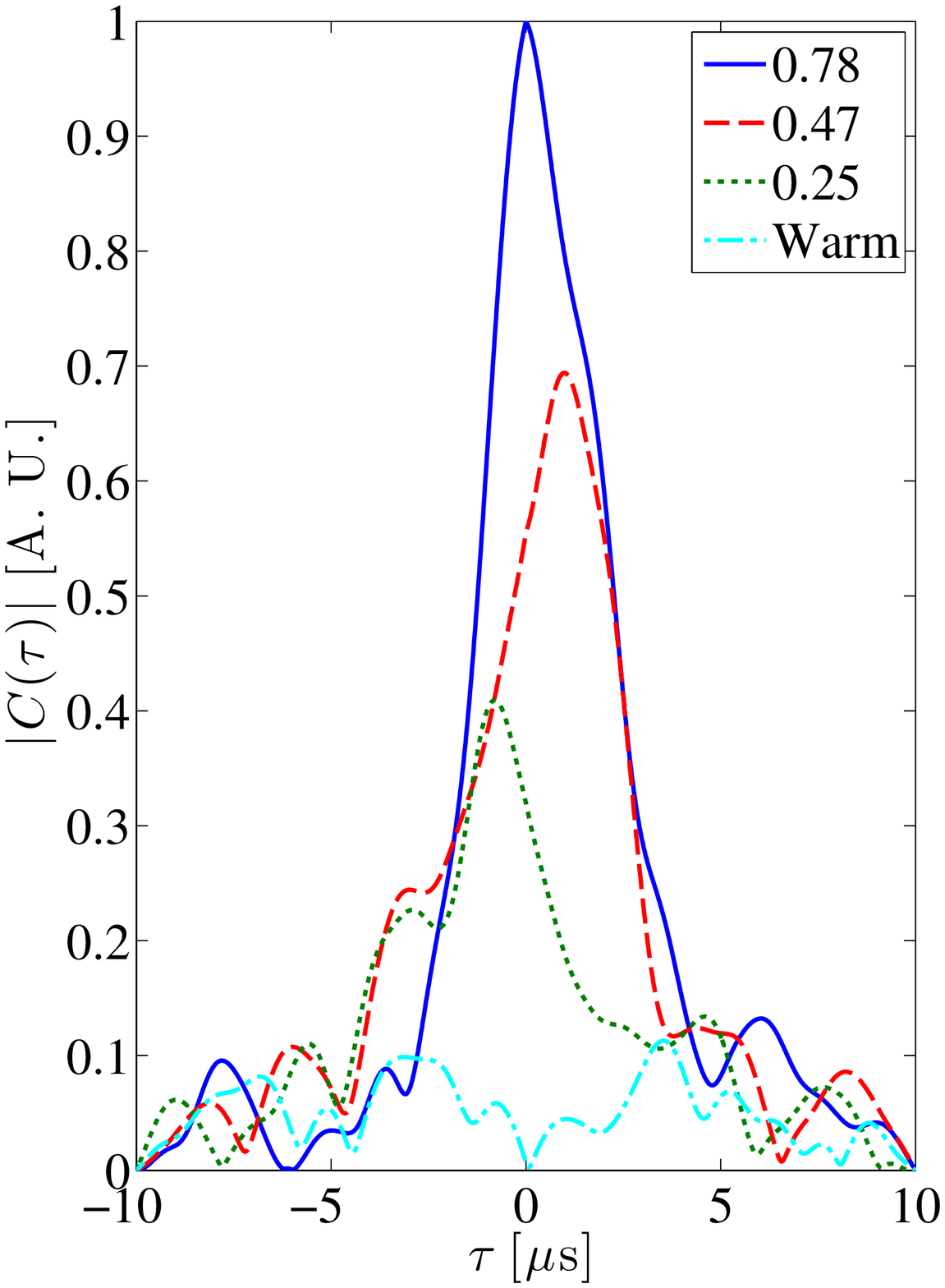}}
\subfigure[]{\raisebox{0.3mm}{\includegraphics[width=0.245\textwidth]{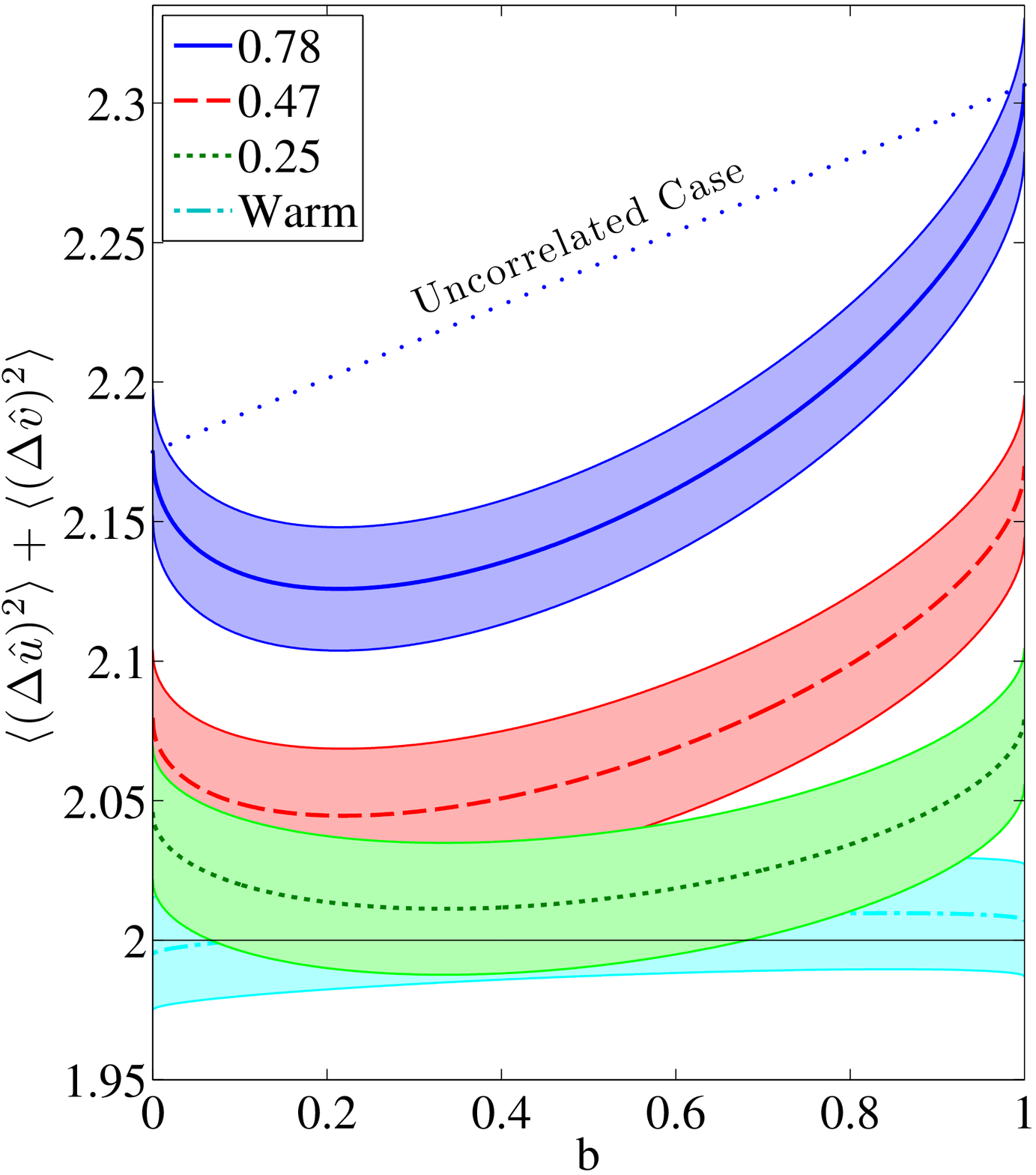}}}
\caption{(Color online) (a) The cross correlation function between the ASE and RASE fields (Eq.~\ref{eq:corrdef}) for three different optical depths 0.25, 0.47, 0.78 and the case where the sample is warmed to 40 K. The cross correlation describes the similarity between the ASE signal and the RASE signal that has been complex conjugated, flipped in time about the $\pi_2$-pulse, and then shifted an amount $\tau$ in time.
(b) The inseparability criterion (Eq.~\ref{eq:crit}) for the same cases as (a) calculated as a function of the $b$ parameter of Eq.~\ref{eq:epr_operators}. The uncorrelated case is shown for $\alpha l = 0.78$. The shaded areas correspond to a confidence interval of 95$\%$ (see \cite{supp})}.
\label{fig:xcorr}
\end{figure}

It can not be seen from time dependent variances and the spectra of Fig.~\ref{fig:RASEvsTime} (b) that the noisy RASE light has coherent correlations with the noisy ASE light. The first way that we characterize these correlations is by evaluating a cross correlation. The rephasing pulse ($\pi_2$) ideally causes a time reversed, conjugated version of the ASE yield, so we define the cross correlation as
\begin{equation}
  C(\tau) = \left\langle \int z_\text{ASE}(t) z_\text{RASE}^*(\tau-t)\, dt\right\rangle \label{eq:corrdef}.
\end{equation}
Here,  $z(t)$ is the complex valued heterodyne signal and the $\tau=0$ has been chosen to correspond with the rephasing $\pi$-pulse ($\pi_2$). The integral is over all time and the $z_\text{ASE}$ and $z_\text{RASE}$ are windowed such that they are zero outside the time windows shown in Fig.~\ref{fig:RASEvsTime}.

Figure~\ref{fig:xcorr}(a) shows  the cross-correlation for  three different optical depths (0.25, 0.47 and 0.78) and for the case where the crystal is warmed to $\sim40$~K. At such warm temperatures, any atomic coherence is lost on nanosecond timescales.  
There is a distinct correlation peak that appears above the warm case for all three cold cases. This clearly confirms a time separated correlation between the ASE and RASE fields and the coherent rephasing effect of $\pi_2$. The temporal width of the correlation peak is broadened by the finite bandwidth of our time separated photon pairs.  The 3.5 $\mu$s width is comparable to the temporal width of the rephasing pulse $\pi_2$ (1.6~$\mu$s) and the bandwidth of the ASE and RASE light (150~kHz). As expected, when the optical depth decreases and the ASE and RASE fields become weaker, the magnitude of this correlation function decreases.

To test the quantum nature of this time separated correlation, we invoke the inseparability criterion for continuous variable states created by Duan et  al. \cite{duan00a}. One can express a maximally entangled state as a co-eigenstate of a pair of EPR type operators 
\begin{equation}
\label{eq:epr_operators}  
\hat{u}~=~\sqrt{b}~\hat{x}_1~+~\sqrt{1 -b}~\hat{x}_2,\quad \hat{v}~=~\sqrt{b}~\hat{p}_1~ -~ \sqrt{1 -b}~\hat{p}_2,
\end{equation}
 where $b \in [0,1]$ is an adjustable parameter describing the weight given to the ASE and RASE fields, $\hat{x}$ and $\hat{p}$ are the light quadratures and the subscript 1 (2) indicates the ASE (RASE) field. For any separable state, the total variance of $\hat{u}$ and $\hat{v}$ satisfies 
\begin{equation}
\label{eq:crit}\langle (\Delta \hat{u})^2\rangle + \langle (\Delta \hat{v})^2\rangle \geq 2.
\end{equation}
For inseparable states, the total variance is bound from below by zero.

By appropriately windowing and then integrating the heterodyne measurement record, values for $\hat{x}$ and $\hat{p}$ for a temporal mode in both the ASE and RASE regions were obtained. Heterodyne detection gives simultaneous and noisy measurements for both quadrature amplitudes \cite{Yuen83}, as opposed to homodyne detection giving good measurements of just one. However as is discussed in \cite{supp}, we can still use Eq.~\ref{eq:crit} as our inseparability criterion. The temporal mode-functions chosen were square and are described in \cite{supp}.

Figure~\ref{fig:xcorr}(b) shows the inseparability criterion versus $b$ for the optical depths of 0.25, 0.47 and 0.78. For $b = 1\,(0)$ the variance is purely the ASE (RASE) field summed over both quadratures.  An indication of how strongly the fields are correlated is the amount the curve dips below the uncorrelated case (straight line between the ASE and RASE values). It is seen that for the $\alpha l = 0.78$ case there is a prominent dip for low $b$. As the optical depth is reduced, and the  ASE and RASE fields weaken this dip becomes less pronounced, but gets closer to the threshold value of 2. The decrease is also seen in the cross correlation measurements. 

\begin{figure}[h!]
{\includegraphics[width=0.45\textwidth ]{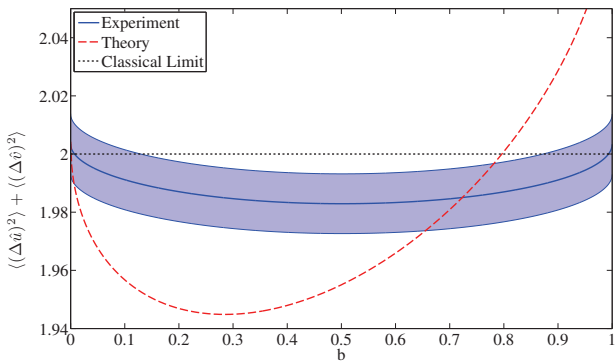}}
\caption{(Color online) Inseparability criterion (Eq.~\ref{eq:crit}) for the ultra-thin case of optical depth 0.046 (solid blue line) calculated as a function of the $b$ parameter (see Eq.~\ref{eq:epr_operators}). The dashed line shows what can be expected theoretically in the ideal case of perfect $\pi$-pulses and no dephasing. The procedure to obtain the theoretical curve is described in \cite{supp}. The band surrounding the experimental curve has a thickness of $1 \sigma$. For $b=0.50$ the confidence level that the experimental  curve is below the below the entanglement threshold is 95.15$\%$}
\label{fig:xcorrThin}
\end{figure}

 Figure~\ref{fig:xcorrThin} shows the inseparability criterion calculated from the measured ASE and RASE fields for a very optically thin sample (0.5 mm, $\alpha l = 0.046$). Here, the inseparability curve dips below the entanglement threshold. At the lowest point in the dip ($b = 0.50$), $\langle (\Delta \hat{u})^2\rangle + \langle (\Delta \hat{v})^2\rangle = 1.983(10) < 2$. The separability criterion is violated to a confidence level of 95.15$\%$. 
 
Shown also in Fig.~\ref{fig:xcorrThin} is the theoretical curve expected from using a heterodyne detector and analytic expressions for the ASE and RASE fields obtained from \cite{ledi10}. The observed dip is smaller than the theoretical prediction but the two are of a comparable size. We attribute the difference to the effects of a finite $T_1$, $T_2$ and imperfect $\pi$-pulses, which are not accounted for in the theory. Even the theoretical dip is small, being limited to 1.94, there are two reasons for this. Firstly at these low optical depths both the ASE and RASE fields are close to the vacuum with on average much less than one photon per temporal mode. This means that even for perfect entanglement the variances of the EPR operators will be both close to one. Secondly,  at such low optical depths, the efficiency of read-out into the RASE field is low and thus the ASE and RASE fields are not perfectly entangled \cite{ledi10}. Most of what is entangled with the ASE light remains in the crystal as an excitation of the ions, rather than coming out as RASE light.  Strategies for overcoming this efficiency problem are discussed below. This inefficiency in the recall of the RASE light is what skews the dip away from $b=0.5$ in the theoretical case. For the theoretical treatment of \cite{ledi10} the only way an atom can relax in the ASE time window is to produce ASE light. This means that theoretically every RASE photon will a have corresponding ASE photon, but due to the inefficiency of the recall the reverse is not true. This is not the case experimentally as atomic relaxation can happen in the ASE window without producing ASE light. This means that unlike the theoretical case there is the possibility of excess noise on the RASE light, and this is a potential reason why the dip is more centered for the experimental curve.

While Fig.~\ref{fig:xcorr} shows multimode correlations between the ASE and RASE fields, the experiment in its current form was unable to demonstrate evidence of multimode entanglement. There were three experimental constraints on the timescales and bandwidth of operation. These were the coherence lifetime $T_2$, the detection recovery time and the available laser power (and therefore the bandwidth of our fixed frequency $\pi$-pulses). These constraints opened a large enough time window to observe evidence of entanglement with just one temporal mode.

The outlook of rephased amplified spontaneous emission as a broadband entangler of photons is promising. Evidence of entanglement was demonstrated in this work but variations of this experiment can be implemented. Changing to a material with a greater coherence time, a detection chain that recovers faster from saturation, and broader bandwidth $\pi$-pulses would all allow multimode demonstrations. \tmyag\ was chosen for its ineffective holeburning feature so as to avoid FID phenomena resulting from small holeburnt features. It also has the benefit of no hyperfine structure at zero magnetic field, which allows for the possibility of broadband operation. In systems with hyperfine structure, which is desired for long term storage \cite{stopped}, FID phenomena can be avoided by phase-matching \cite{Damon2011} or by a four level echo scheme \cite{beav11}. 
Operation at telecommunication wavelengths is possible using erbium dopants where the potential for long lived spin coherences has recently been identified \cite{mcau12}.  
The entanglement evidence seen here is small and on the edge of statistical significance, even in the ideal two level atom case RASE does not achieve perfect entanglement. This can be seen as a consequence of a fixed optical depth. Ideally, weak interaction with the light is required when generating the ASE, this means that there is on average a small number of photons per spatio-temporal mode. Then, strong interaction with the light is required to efficiently retrieve the RASE. Beavan et al.'s four level scheme \cite{beav11}, Raman transitions with changing coupling strengths, and Q-switched cavities \cite{inprep} are all ways of achieving changes in the effective optical depth. 

In conclusion we have demonstrated the generation of ASE and RASE fields from a cryogenically cooled \tmyag\ crystal for different optical depths. The effect of the second $\pi$-pulse ($\pi_2$) is clearly seen in the spectrum of the optically thick case. Clear time separated correlation peaks are observed between the ASE and RASE fields for optical depths ranging from 0.25 to 0.78 demonstrating coherent rephasing of the ASE field.  For an optical depth of 0.046 we see evidence, at the $95\%$ confidence level, that the ASE and RASE fields are entangled. 

We note that a similar work demonstrating time separated correlations in a praseodymium doped crystal has been performed independently \cite{beav12}.

We acknowledge that this research was funded by the New Zealand Foundation for Research Science and Technology under Contract No. NERF-UOOX0703.


\end{document}